\documentclass[aps,prd,twocolumn,floatfix,nofootinbib,superscriptaddress,tightenlines]{revtex4}

\usepackage[dvipsnames]{xcolor}
\usepackage{amsmath}
\usepackage{dcolumn}
\usepackage{lipsum}
\usepackage{amssymb}
\usepackage{epsfig}
\usepackage{graphics}
\usepackage{graphicx}
\usepackage{subcaption}
\usepackage[justification=raggedright,singlelinecheck=false]{caption}

\usepackage{url}
\usepackage{epsfig}
\usepackage{graphicx}
\usepackage{amsmath}
\usepackage{bm}
\usepackage{setspace}
\usepackage{lscape}
\usepackage{amsthm}
\usepackage{bbold}
\usepackage{dcolumn}
\usepackage{epsfig}
\usepackage{graphics}
\usepackage{graphicx}
\usepackage{mathrsfs}

\usepackage{bm}
\usepackage{xspace}
\usepackage{cancel}
\usepackage{float}
\usepackage{multirow}
\definecolor{darkgreen}{rgb}{0,0.5,0}
\definecolor{purple}{rgb}{0.5,0,0.5}
\definecolor{nblue}{rgb}{0.0,0.0,0.50}
\definecolor{scarlet}{rgb}{1.0,0.2,0}
\definecolor{darkmagenta}{rgb}{0.55, 0.0, 0.55}
\definecolor{darkolivegreen}{rgb}{0.33, 0.42, 0.18}
\definecolor{darkcandyapplered}{rgb}{0.64, 0.0, 0.0}

\usepackage[colorlinks=true, pdfstartview=FitV, linkcolor=purple, citecolor= purple, urlcolor=blue]{hyperref}

\newcommand{\be}{\begin{equation}}

\newcommand{\ee}{\end{equation}}
\newcommand{\bea}{\begin{eqnarray}}
\newcommand{\eea}{\end{eqnarray}}
\newcommand{\beas}{\begin{eqnarray*}}
\newcommand{\eeas}{\end{eqnarray*}}
\newcommand{\nn}{\nonumber}
\newcommand{\slsh}[1]{{\not \! #1}}

\begin{document}
\title{\textbf{Dynamical Mass Generation in QED: Miransky scaling and Schrödinger-like infinite well and barrier potentials supporting a bound state}}

\author{Juan D. García-Muñoz}
\email{juan.garciam@cinvestav.mx}
\affiliation{Mesoamerican Centre for Theoretical Physics,
Universidad Aut\'onoma de Chiapas, Carretera Zapata Km.~4, Real
del Bosque (Ter\'an), Tuxtla Guti\'errez, Chiapas 29040, M\'exico}
\affiliation{Departamento de Física, Cinvestav, Av. Instituto Politécnico Nacional 2508, 07360 Ciudad de México, México}

\author{A. Alfaro}
\email{alfaro@unach.mx}
\affiliation{Facultad de Ciencias en Física y Matemáticas, Universidad Aut\'onoma de Chiapas, 
Carretera Emiliano Zapata Km.~8, Rancho San Francisco, Ciudad Universitaria Ter\'an, Tuxtla Guti\'errez, Chiapas 29040, M\'exico}

\author{L. X. Guti\'errez-Guerrero}
\email{lxgutierrez@conacyt.mx}
\affiliation{CONACyT-Mesoamerican Centre for Theoretical Physics,
Universidad Aut\'onoma de Chiapas, Carretera Zapata Km.~4, Real
del Bosque (Ter\'an), Tuxtla Guti\'errez, Chiapas 29040, M\'exico}

\author{A. Raya}
\email{alfredo.raya@umich.mx}
\affiliation{Facultad de Ingeniería Eléctrica, Universidad
Michoacana de San Nicol\'as de Hidalgo, Edificio $\Omega-2$, Ciudad
Universitaria, Morelia, Michoac\'an 58040, M\'exico}
\affiliation{Centro de Ciencias Exactas, Universidad del B\'{\i}o-B\'{\i}o,\\ Avda. Andr\'es Bello 720, Casilla 447, 3800708, Chill\'an, Chile.}

\begin{abstract}
 In this study, we revisit the Schwinger–Dyson equation for the electron propagator in QED in three- and four-space--time dimensions. Our analysis addresses the non-perturbative phenomenon of dynamical chiral symmetry breaking which demands a critical value of the coupling for the dynamical generation of electron masses, encoded in the infrared behavior of the said Green function. With a minimalistic truncation of the infinite tower of equations and adopting standard assumptions, the resulting gap equation is linearized and transformed into a Schrödinger-like equation with an auxiliary potential barrier (well) subjected to boundary conditions for both high and low momenta. Then, the dynamical mass is associated with the zero mode of the corresponding Schrödinger-like operator and adheres to the Miransky scaling law, as expected. 
\end{abstract}
\maketitle
\section{Introduction}
The study of dynamical mass generation in gauge theories has a rich history, deeply rooted in the exploration of non-perturbative phenomena in quantum field theory~\cite{MiranskyBook}. Early motivations arose from the pre-quantum chromodynamics (QCD) era, with 
Nambu-Jona-Lasinio (NJL) model, which provided a foundational framework for understanding mass generation through spontaneous symmetry breaking~\cite{NJL1,NJL2,klevansky,BUBALLA2005205,Volkov:2005kw}. However, the exploration of gauge theories, particularly those with strong coupling, revealed more intricate mechanisms, such as confinement and chiral symmetry breaking~\cite{Miransky:1985gbh}.
This understanding was later extended to technicolor models, which sought to explain electroweak symmetry breaking through a new type of strong interactions, responsible for dynamically generating masses for the $W$ and $Z$ bosons without elementary Higgs fields~\cite{WeinbergTC,SusskindTC,HoldomTC,AppelTC1,AppelTC2,AppelTC3,HILL2003235}.
Parallel developments in quantum electrodynamics (QED) and its lower-dimensional counterpart, QED$_3$, further illuminated the role of dynamical mass generation for fermions (electrons). In QED, the phenomenon was studied in the context of the Schwinger mechanism and the critical coupling for chiral symmetry breaking~\cite{Baker,Fomin:1976af,Miransky:1985gbh,Miransky1985,Curtis:1990zs,Curtis:1990zr,Curtis:1991fb,Curtis:1992gm,Curtis:1992jg,Curtis:1993py,Atkinson:1993mz,Bashir:1994az,Bashir:1995qr,Kizilersu:1995iz,Bashir:1997qt,GusyninDR,Kizilersu:2009kg,Kizilersu:2014ela}, whereas in QED$_3$, the dynamical generation of fermion (or electron, which we use indistinctly throughout the manuscript) mass  has raised interest on its own for several decades of study~\cite{Appel1,Appel2,Nash,Dagotto:1988id,Dagotto:1989td,Pennington:1990bx,Burden:1991uh,Curtis:1992gm,Maris:1995ns,PhysRevD.54.4049,Eichmann:2016yit,Fischer:2004nq,Bashir:2004yt,Bashir:2005wt,Bashir:2008fk,Bashir:2009fv,Gusynin16}. QED$_3$ became a focal point in condensed matter physics due to its relevance to high-temperature superconductivity~\cite{DOREY1990107,Franz01,Herbut02} and to graphene more recently, within the so-called pseudo or reduced QED~\cite{MARINO1993551,GONZALEZ1994595,Gorbar2001,Herbut2007,Herbut2009,Gamayun2010,Teber2012,Alves2013,Nascimento2015,kotikov2016,Carrington2019,Carrington:2020qfz,Cuevas2020,Casimiro2020,Carrington:2022wjj,Alvino2022,Casimiro,Marino2024} which unlike the logarithmic static interaction between charges of QED$_3$, it generates a Coulomb-like static charge interaction by intertwining the dynamics of fermion and gauge field propagation in mixed dimensions. Naturally, one cannot overstate the crucial role played by dynamical chiral symmetry breaking and confinement as emergent phenomena that shape the hadron spectrum. This framework has reached a remarkable level of maturity as a field in its own right, as has been reviewed, for example, in~\cite{Roberts:1994dr, Pennington:2005be,Fischer:2006ub, Eichmann:2016yit, Qin:2020jig, Ding:2022ows, Raya:2024ejx}. A benchmark advancement came with the formulation of the Miransky scaling law~\cite{Miransky1985,Miransky:1996pd}, which describes the non-perturbative behavior of the mass gap near critical coupling and has provided a framework for understanding the phase structure of gauge theories. This scaling law has been instrumental in elucidating the interplay between confinement, chiral symmetry breaking, and dynamical mass generation, offering insights into both particle physics and condensed matter systems. 

The Miransky scaling law arises naturally when addressing non-perturbative dynamics within QED~\cite{Miransky1985,Miransky:1996pd}, a weakly-coupled theory. A favorite starting point is the electron gap equation and the assumption that the fine structure constant might grow freely beyond its well-known physical value 1/137. Assuming massless electrons to start with, so long as $\alpha$ is weak, they remain massless, and the theory is scale-invariant and symmetric under chiral transformations. However, when the coupling exceeds a critical value $\alpha_c$, these electrons acquire a  mass through self-interactions, thus breaking chiral symmetry. The stronger the coupling, the larger the dynamical mass. The essentially-singular behavior of the mass on the strength of the coupling reflects the non-analytic nature of the phase transition~\cite{Miransky:1996pd}.
The scaling law provides insights into the behavior of theories near critical points, like lower dimensional versions of QED and other strongly coupled gauge theories.

The starting point within this framework is an educated  truncation of the corresponding Schwinger-Dyson equations (SDEs), often modeling the non-perturbative form of the fermion-photon vertex and including vacuum polarization effects. Rather than pursuing a state-of-the-art truncation, we choose to work with a minimalistic raibow-ladder inspired truncation that exhibits the general features of the phenomenon at the time that permits an analytic treatment of the interaction kernel.  These assumptions have enabled the exploration of dynamical mass generation, chiral symmetry breaking, and hadron properties in QED and QCD~\cite{Roberts:1994dr, Pennington:2005be,Fischer:2006ub, Eichmann:2016yit, Qin:2020jig, Ding:2022ows, Raya:2024ejx}. While this simplified kernel has limitations, ongoing research continues to refine and extend this approach, enhancing its applicability and accuracy in various physical contexts. In QCD, the omission of gluon self-interactions and vertex corrections leads to inaccuracies in the description of some hadron observables. To address these limitations, extensions of the rainbow-ladder truncation have achieved a prominent level in the prediction of hadron properties, competitive with lattice simulations~\cite{Roberts:1994dr, Pennington:2005be,Fischer:2006ub, Eichmann:2016yit, Qin:2020jig, Ding:2022ows, Raya:2024ejx}. 

In this article, we revisit the gap equation  in QED in three- and four-space--time dimensions within a minimalistic rainbow-ladder inspired truncation. Upon a linearization of the resulting integral equation, we map it onto a differential equation with Schrödinger-like form in an auxiliary potential barrier (well) that supports a zero-energy bound state when the coupling parameter reaches a critical value consistent with other truncations and which must fulfill boundary conditions at large and low momentum. These boundary conditions impose a quantization rule that allows to identify such mode with a mass that fulfills the Miransky scaling law.  
For this purpose, we organize the remaining of this article as follows:
In Section~\ref{SDE}, we establish the SDE for QED. We analyze these equations for QED$_3$ including vacuum polarization effects in subsection~\ref{QED3} and within a quenched approximation for QED$_4$
  in Section~\ref{QED4}. In both cases, we study the corresponding Miransky scaling laws that emerge from linearizing the gap equation. In Section~\ref{S-EQ}, we study the gap equation in the form of a Schrödinger-like equation with an effective potential and study the bound state problem. Finally, in Section~\ref{Summary}, we provide a summary and outlook of the present work.

\section{Schwinger-Dyson equations}
\label{SDE}
\begin{figure}[t!]
   \vspace{-2cm}
   \centering
    \includegraphics[width=0.7\columnwidth,angle=-90]{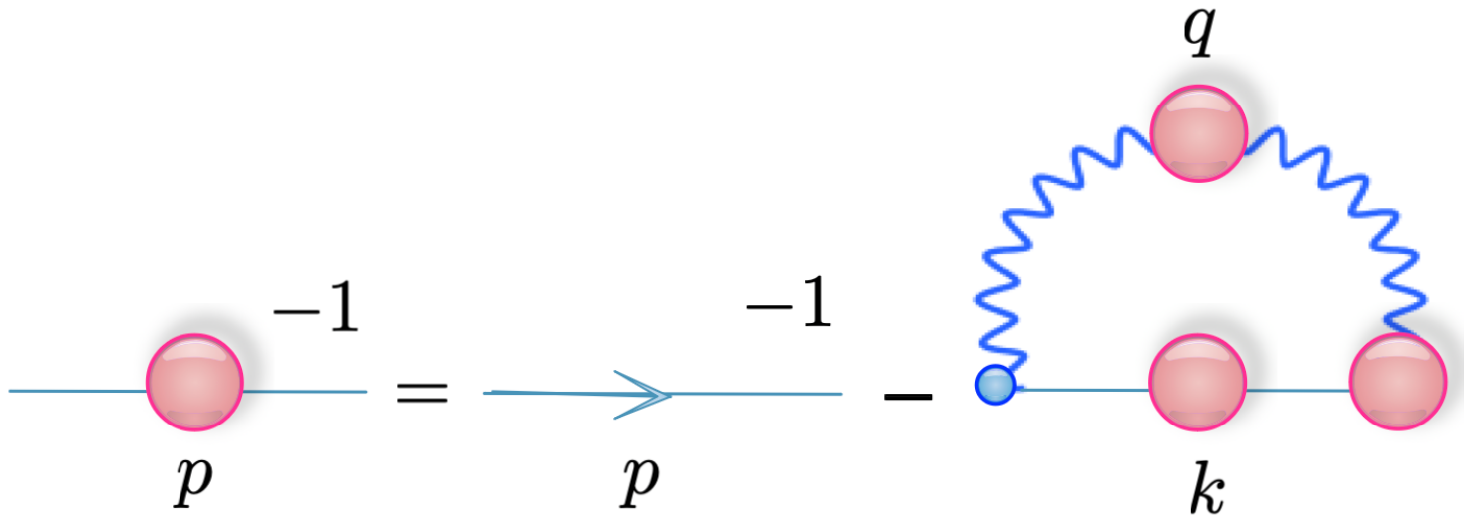}
    \vspace{-2cm}
    \caption{Diagrammatic representation of the fermion gap equation in  QED. The fermion and photon propagators are represented by solid lines and wavy lines, respectively, while fully dressed quantities are indicated by filled magenta circles.}
    \label{fig:sde}
\end{figure}

SDEs in quantum field theory represent an infinite set of dynamic relationships among its Green functions. To solve these equations, a suitable truncation method is necessary. Beyond perturbation theory, the only known systematic truncation scheme requires the modeling of the involved Green functions and interaction kernels to account for non-perturbative phenomena. 
To explore the dynamic generation of fermion masses, we begin with the fermion (quark or electron) gap equation and model the interaction kernel. In QED, a diagrammatic representation of the gap equations for the electron propagator can be seen in Fig.~\ref{fig:sde},
which corresponds to the following equation in $d$-dimensions~\cite{Olivares:2021svj},
\begin{eqnarray}
	S^{-1}(p)&=&S_{0}^{-1}(p)\nonumber\\
	&&\hspace{-10mm}+4\pi i \alpha\int\frac{d^{d}k}{(2\pi)^{d}}\Gamma^{\mu}(k,p)S(k)\gamma^{\nu}\Delta_{\mu\nu}(q),
	\label{sd}
	\end{eqnarray} 
where $\alpha=e^2/(4\pi)$, $q=k-p$, $e$ is the electromagnetic coupling and we use the most general form of the propagator as $S(p) = {F(p^2)}/{(\slsh{p}-\Sigma(p^2))}$, $\Sigma(p^2)$ and  $F(p^2)$, are the well-known mass function and wavefunction renormalization, respectively. The full photon propagator $\Delta_{\mu \nu}(q)$ is
\bea
\Delta_{\mu \nu}(q) = -\frac{G(q^2)}{q^2}
\left( g_{\mu \nu} - \frac{q_{\mu} q_{\nu}}{q^2} \right)
- \xi \frac{q_{\mu} q_{\nu}}{q^4}  \;,
\eea
where $\xi$ is the covariant gauge parameter, with $\xi=0$ corresponding to the Landau gauge, and $G(q^2)$ is the photon wavefunction renormalization function, which at tree level in perturbation theory, corresponds to $G^{(0)}(q^2)=1$. Additionally, $\Gamma^\mu(k,p)$ is the full fermion-photon vertex.
equation \ref{sd} is a matrix equation that can be cast into a system of two scalar non-linear integral equations, 
\bea
\frac{1}{F(p^2)}&=& 1-\frac{4\pi i \alpha}{d\ p^2}\int \frac{d^dk}{(2\pi)^d} {\rm Tr}[{\not \! p}\gamma^\mu S(k) \Gamma^\nu(k,p)]\Delta_{\mu\nu}(q)\nn\\
\frac{\Sigma(p^2)}{F(p^2)}&=&\frac{4\pi i \alpha}{d}\int \frac{d^dk}{(2\pi)^d}{\rm Tr}[\gamma^\mu S(k) \Gamma^\nu(k,p)]\Delta_{\mu\nu}(q)\;.\label{eq:rainbow}
\eea
In the following, we analyze these equations in both three- and four-space--time dimensions with suitable forms of the photon propagator that lead to a Miransky scaling of the dynamical mass.
\subsection{QED in Three Dimensions:}
\label{QED3}

\begin{figure}[t!]
   \vspace{0cm}
   \centering
\includegraphics[width=0.7\columnwidth,angle=-90]{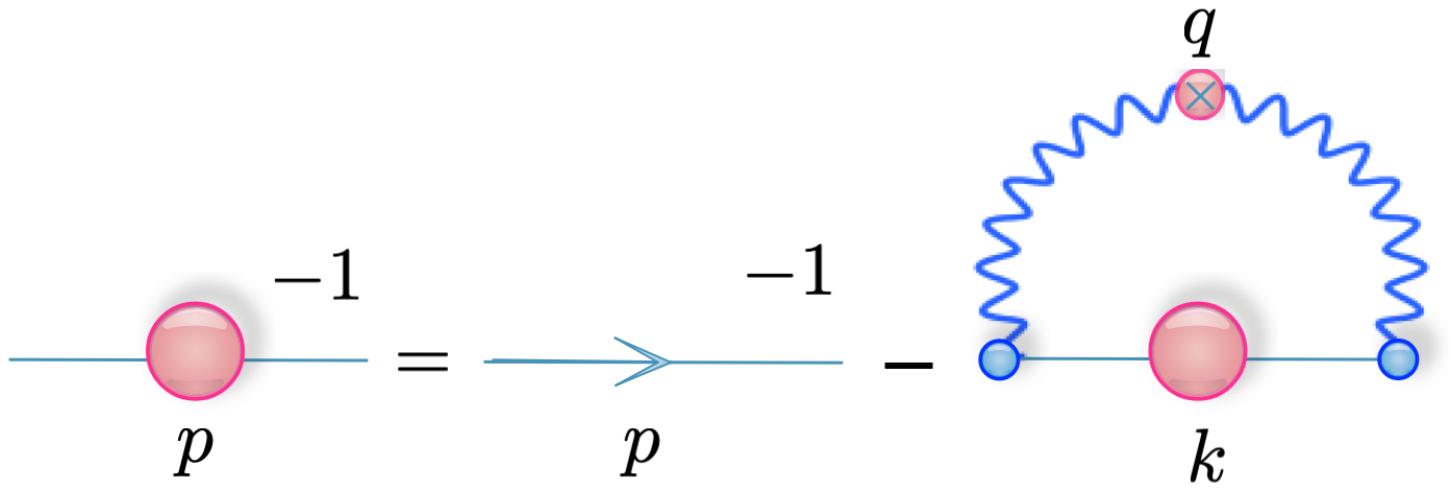}
    \vspace{-0.5cm}
\caption{Diagrammatic representation of the electron gap equation in QED3 including vacuum polarization effects at the leading order of the $1/N_f$ approximation. The fermion and photon propagators are represented by solid lines and wavy lines, respectively. The full electron propagator is represented by the filled magenta circle, whereas the vacuum contribution in the photon propagator is indicated by the cross on top of the small magenta circle.}
\label{fig:qed3}
\end{figure}

We begin by considering the case of QED in three-space--time dimensions, i.e., $d=3$, where the coupling constant  $e^2$ has dimensions of mass, which implies that all scales in the theory can be expressed in terms of this coupling. Consequently, there is no need to regulate integrals, as the theory becomes super-renormalizable. Additionally, there is no critical value of  $e^2$ needed to dynamically break chiral symmetry. If chiral symmetry is broken at a particular value of the coupling, it will be broken for any arbitrary value of it. Nevertheless, by including loops of massless fermions, as shown in Fig.~\ref{fig:qed3}, a new coupling parameter can be defined as ${\rm e}^2=e^2N_f$ where $N_f$ is the number of fermion families circulating on these loops. 

As $N_f\to\infty$ while retaining $\rm e^2$ fixed, in Ref.~\cite{Appel1} it was first noticed that there exists a critical number $N_f^c$ of electron families above which chiral symmetry is restored. The argument goes as follows: In the leading order of the $1/N_f$ expansion, the ingredients of the gap equation are the bare vertex $\Gamma^\nu\to \gamma^\nu+{\cal O}(1/N_f)$ and a trivial behavior of the wavefunction renormalization, $F(p)~\simeq~1+{\cal O}(1/N_f)$. Then, summing up bubbles of electron loops, the gap equation becomes
%
\be
\Sigma(p)=\frac{{\rm e}^2}{2\pi^2 N_f p}\int_{0}^{\infty}dk\frac{k\Sigma(k)}{k^2+\Sigma^2(k)}\ln\left| \frac{{\rm e}^2/8+8|k+p|}{{\rm e}^2/8+8|k-p|}\right|.~\label{gapnlqed3}
\ee
 It is immediate to note that there exists a trivial solution $\Sigma(p) = 0$, which follows directly from the fact that in perturbation theory, a massless particle cannot acquire mass. A non-trivial analytic solution of the above equation~\eqref{gapnlqed3} is not possible to obtain due to the non-separable nature of the interaction kernel. However, let us perform the following simplifications. Expanding the logarithm for $k\gg p$ and $k\ll p$, we get
\bea
\Sigma(p)&=&\frac{{\rm e}^2}{\pi^2 N_f p}\int_{0}^{p}dk\frac{k\Sigma(k)}{k^2+\Sigma^2(k)}\left(\frac{k}{{\rm e}^2/8+p} \right)\nonumber\\
&&+\frac{{\rm e}^2}{\pi^2 N_f p}\int_{p}^{\infty}dk\frac{k\Sigma(k)}{k^2+\Sigma^2(k)}\left(\frac{p}{{\rm e}^2/8+k} \right) .
\eea
For ${\rm e}^2\gg p$, upon neglecting terms of $\Sigma^2(p)$, the above equation can be expressed as the differential equation
\be 
\frac{d}{dp}\left[p^2 \frac{d\Sigma(p)}{dp} \right]=-\frac{8}{\pi^2 N_f} \Sigma(p), \label{gap3}
\ee
subjected to the boundary conditions
\be
0\le \Sigma(0) \le \infty, \qquad \left[p\frac{d\Sigma(p)}{dp}+\Sigma(p)\right]\Bigg|_{p={\rm e^2}}=0. \label{BC3}
\ee
This conditions, and all those written in this section, are chosen to guarantee the regularity of solutions at zero momentum and the correct asymptotic behavior at large momentum, consistent with the underlying QED dynamics. Moreover, some of them are directly calculated from their corresponding integral equation. Seeking for a power law solution,
\be
\Sigma(p)=p^s, \label{sol3}
\ee
we obtain that
\be
s_\pm = -\frac12 \pm \frac12 \sqrt{1-\frac{32}{\pi^2N_f}},
\ee
and hence we notice that there exists a critical value $N_f^c=32/\pi^2$ that distinguishes an oscillatory behavior from a power-law solution. Boundary conditions demand the former behavior, which naturally leads to the Miransky scaling law for the dynamical mass,

\begin{figure}[t!]
   \vspace{0cm}
   \centering
\includegraphics[width=0.7\columnwidth,angle=-90]{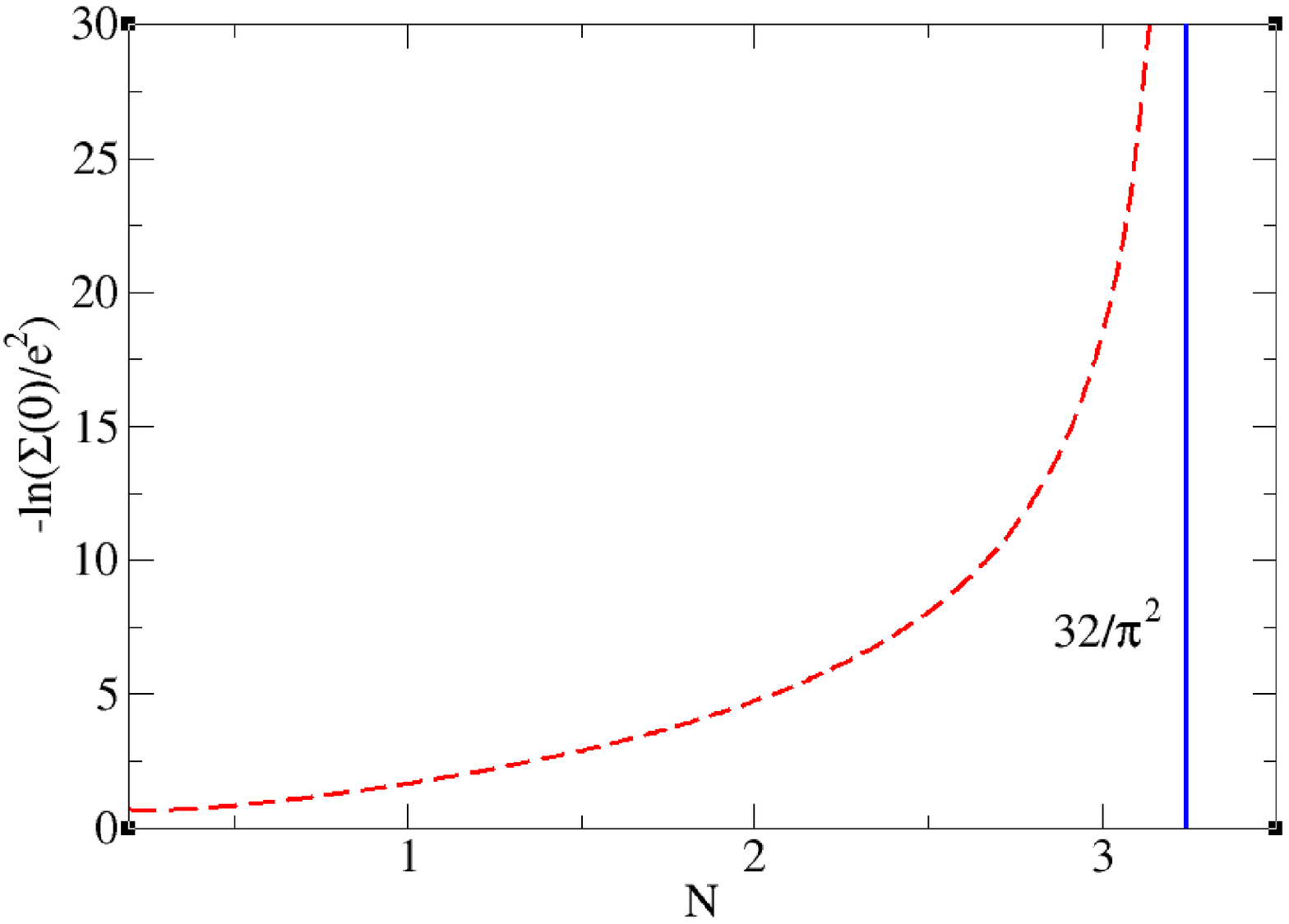}
    \vspace{-0.5cm}
\caption{ Plot the logarithm of Equation (\ref{qed3scaling}) as a function of   
$N_f$. The curve provides indications of a singularity as $N_f$ approaches to
$N_f^c=32/\pi^2$.
  }
\label{fig:qedq}
\end{figure}

\be \label{qed3scaling}
\Sigma(0)={\rm e}^2 \exp\Bigg[\frac{-2\pi}{\sqrt{\frac{N_f^c}{N_f}-1}} \Bigg].
\ee
These analytical results are derived from the linearization of the original problem,
which is actually exact within bifurcation theory,  see Refs. \cite{Pennington:1990bx,Bashir:2008fk,Gusynin:1995bb,Maris:1995ns,Gusynin:1998kz}.  
This linearization of the gap equation provides an exact determination of the critical coupling, simplifies the analysis without drastic approximations, guarantees a multiplicatively renormalizable mass function, and naturally yields Miransky scaling together with the critical coupling and the critical number of flavors.\\
Fig.~\ref{fig:qedq} illustrates the behavior of $\Sigma(0)$ as a function of $N_f$, showing that the curve tends asymptotically toward $32/\pi^2$.
The bifurcation analysis~\cite{Miransky:1985gbh,Fomin:1976af} carried out on the original non-linear equation~\eqref{gapnlqed3} reveals that beyond a critical value of the coupling constant, a non-trivial solution $\Sigma(p) \ne 0$ {\em bifurcates} away from the trivial one. In other words, mass generation occurs dynamically even when starting from massless electrons. 

Based on the preceding discussion, we conclude that the dynamical mass in QED$_3$
  exhibits scaling behavior with respect to the number of fermion flavors \cite{Appelquist:1988sr}. 
What follows is the analysis in four dimensions.


\subsection{QED in Four Dimensions:}

\begin{figure}[b!]
   \vspace{-2cm}
   \centering
    \includegraphics[width=0.7\columnwidth,angle=-90]{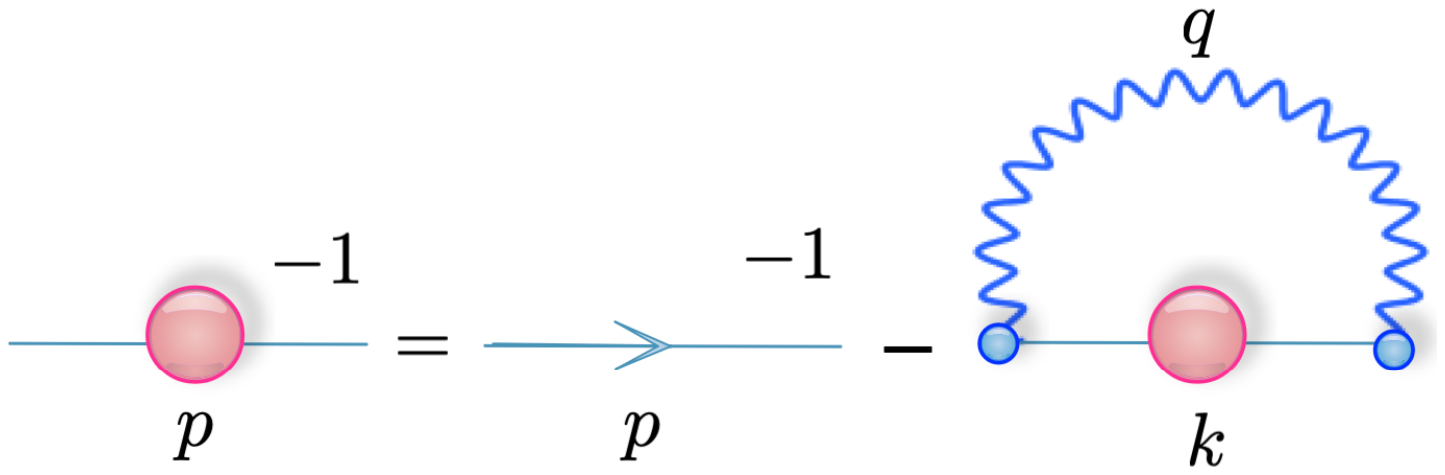}
    \vspace{-2cm}
    \caption{SDE for the fermion propagator in rainbow-ladder approximation, where the following ans\"atze are considered $\Gamma^\mu(k,p) \to \gamma^\mu$ and $\Delta_{\mu\nu}(q) \to \Delta_{\mu\nu}^{(0)}(q)$. }
    \label{fig:sde2}
\end{figure}

\label{QED4}
In the QED$_4$ case, we follow a similar approach. Adopting the rainbow-ladder truncation depicted in Fig.~\ref{fig:sde2}, which amount to consider the bare vertex and photon propagator in the interaction kernel, working in Landau gauge we have $F(p^2)=1$ and thus we have the following expression for the mass function,
\bea \label{Meq}
\Sigma(p^2)&=&\frac{3\alpha}{4\pi}\Bigg[ \int_{0}^{p^2} dk^2 \frac{k^2}{p^2} \frac{\Sigma(k^2)}{k^2+\Sigma^{2}(k^2)}\nonumber \\
&&+\int_{p^2}^{\Lambda^{2}}dk^2 \frac{\Sigma(k^2)}{k^2+\Sigma^{2}(k^2)}\Bigg].
\eea
\\
\\
Given the complexity of attempting to solve equation~\eqref{Meq} analytically, 
and exploiting the advantage of the rainbow-ladder approximation,  
we proceed to linearize it in the spirit of bifurcation theory~\cite{Miransky:1985gbh,Fomin:1976af}.
 We first perform the change of variables $x=p^2$ and $y=k^2$. 
Furthermore, by demanding than the non-linear terms $\Sigma(k)=0$ and adding an infrared cut-off $\kappa$ such that $\Sigma(\kappa)=\kappa$ to preserve the scale non-invariance of the integral equation~\eqref{Meq}, we end up with
\begin{equation} \label{hE1}
    \Sigma(x)=\frac{3\alpha}{4\pi}\left[\frac{1}{x}\int_{\kappa^2}^{x}dy \frac{y\Sigma(y)}{y}+\int_{x}^{\Lambda^{2}}dy \frac{\Sigma(y)}{y}\right].
\end{equation}
The above linear integral equation~\eqref{hE1} transforms into the differential equation
\begin{equation}\label{hE2}
    x^{2}\Sigma ''(x)+2x\Sigma '(x)+\frac{\hat{\alpha}}{4}\Sigma(x) = 0,
\end{equation}
subjected to the boundary conditions (BC)
\begin{equation}\label{hE3}
    \frac{d}{dx}\left(x\Sigma(x)\right)\bigg|_{x=\Lambda^{2}}=0,\quad \frac{d}{dx}\Sigma(x)\bigg|_{x=\kappa^2}=0,
\end{equation}
with $\hat{\alpha}=3\alpha/\pi$.
Seeking a solution to~\eqref{hE2} of the form $\Sigma(x)=x^s$, we straightforwardly find
\begin{equation}\label{hE4}
s_\pm=-\frac12 \pm \frac{\sqrt{1-\hat{\alpha}}}{2}.
\end{equation}
Thus, depending on whether $\hat{\alpha}>1$ or $\hat{\alpha}<1$ we can find real or complex values for $s_\pm$. BC demand $\hat{\alpha}>1$, and thus
the general solution to \eqref{hE2} can be expressed as
\begin{equation}\label{hE5}
\Sigma(x)=C_+ x^{-\frac{1}{2}+i\frac{\hat \tau}{2}}+C_- x^{-\frac{1}{2}-i\frac{\hat \tau}{2}},
\end{equation}
with $\hat{\tau}=\sqrt{\hat{\alpha}-1}\in \mathbb{R}$. Inserting back this solution into the BC, we have that
\begin{equation}
\left( \frac{\Lambda}{\kappa}\right)^{2i\hat{\tau}}= \frac{(\frac{1}{2}-i\frac{\hat \tau}{2})^2}{(\frac{1}{2}+i\frac{\hat \tau}{2})^2}.\label{hardlin}
\end{equation}
For $\hat\alpha\to 1$,
\begin{equation}\label{hE7}
\ln\left( \frac{\Lambda}{\kappa}\right)=\frac{k \pi}{\hat \tau}-2, \qquad k\in \mathbb{N}.
\end{equation}
For the minimal energy state, $k=1$, we find
\begin{equation}\label{hE8}
\frac{\Lambda}{\kappa}=\exp \left( \frac{\pi}{\hat \tau}-2\right),
\end{equation}
which again corresponds to a Miransky scaling~\cite{Miransky:1985gbh,Miransky:1996pd}.

To soften the linearization of the original equation~\eqref{Meq}, we  set $m^{2}=\Sigma^{2}(0)$ in the denominator of this expression, 
\begin{equation} \label{E1}
    \Sigma(x)=\frac{3\alpha}{4\pi}\left[\frac{1}{x}\int_{0}^{x}dy \frac{y\Sigma(y)}{y+m^{2}}+\int_{x}^{\Lambda^{2}}dy \frac{\Sigma(y)}{y+m^{2}}\right].
\end{equation}
avoiding the need to include an IR cut-off to the integrals.
The above equation~\eqref{E1} can be transformed into the following differential equation
\begin{equation}\label{E2}
    x^{2}\Sigma ''(x)+2x\Sigma '(x)+\frac{3\alpha}{4\pi}\frac{x}{x+m^{2}}\Sigma(x) = 0,
\end{equation}
satisfying the BC
\begin{equation}\label{E3}
    \frac{d}{dx}\left(x\Sigma(x)\right)\bigg|_{x=\Lambda^{2}}=0,\quad \Sigma(x)\bigg|_{x=0}=m,
\end{equation}
dubbed  as the UV and IR conditions, respectively. Introducing the additional change of variable $z = -x/m^{2}$, we arrive at one particular case of the hypergeometric differential equation 
\begin{equation} \label{E4}
    z(1-z)\Sigma ''(z)+2(1-z)\Sigma' (z)-\frac{\alpha}{4\alpha_{c}}\Sigma(z)=0,
\end{equation}
where $\alpha_c=\pi/3$. The general solutions to the previous equation are given by
\begin{equation}\label{solE}
    \Sigma(z) =m F\left(\frac{1}{2}\pm\sigma,\frac{1}{2}\mp\sigma;2;z\right),
\end{equation}
where $F(a,b;c;z)$ is the hypergeometric function and $\sigma=(1/2)\sqrt{1-\alpha/\alpha_c}$. Note that both solutions are the same since $F(a,b;c;z)=F(b,a;c;z)$. The behavior of the generated mass solutions $\Sigma(z)$ has been analyzed in detail in Ref.~\cite{Bloch}. 
The discussion goes as follows. Imposing the UV condition to the solution $\Sigma(z)$, which states that, as $\tau\to 0$,
\begin{equation*}
     \left(\frac{\Lambda^2}{m^2}\right)^{i\tau}=-\frac{\Gamma(-i\tau)\Gamma^2(\frac{1}{2}+\frac{i}{2}\tau)}{\Gamma(i\tau)\Gamma^2(\frac{1}{2}-\frac{i}{2}\tau)}.
\end{equation*}
Using $\overline{\Gamma(z)}=\Gamma(\overline{z})$ and defining $\Gamma(i\tau)\equiv r_1\exp(i\theta_1)$ and $\Gamma(\frac{1}{2}+\frac{i}{2}\tau)\equiv r_2\exp(i\theta_2)$, respectively,  we get
\begin{equation}
    \left(\frac{\Lambda^2}{m^2}\right)^{i\tau}=\exp(i\theta),
    \label{sk3}
\end{equation}
where $\theta=\pi-2\theta_1+4\theta_2$.
As we want $\theta$ for $\tau\ll 1$, upon Taylor expanding the gamma functions,
\begin{equation*}
    \Gamma(i\tau)\approx\frac{1}{i\tau}(1+i\tau\psi(1))=\psi(1)-\frac{i}{\tau},
\end{equation*}
and,
\begin{equation*}
    \Gamma\left(\frac{1}{2}+\frac{i}{2}\tau\right)\approx\Gamma\left(\frac{1}{2}\right)\left(1+\frac{i}{2}\tau\psi\left(\frac{1}{2}\right)\right),
\end{equation*}
where $\psi(x)$ represents the digamma function, we get,
\begin{eqnarray*}
    \theta_1&=&{\rm arg}\left(-\gamma-\frac{i}{2}\tau\right)
    \approx\frac{\pi}{2}-\gamma\tau,\\
    \theta_2&=&{\rm arg}\left(\Gamma\left(\frac{1}{2}+\frac{i}{2}\tau\right)\right)
    \approx\frac{\tau}{2}\psi\left(\frac{1}{2}\right).
\end{eqnarray*}
Furthermore, using $\psi\left(\frac{1}{2}\right)=-\gamma-2\ln(2)$ we finally get,
\begin{equation*}
    \theta\approx-4\tau\ln(2).
\end{equation*}
Substituting the above result into equation~\eqref{sk3}, we find,
\begin{equation*}
    \ln\left(\frac{\Lambda}{m}\right)=\frac{k\pi}{\tau}-2\ln(2), \qquad k\in {\mathbb N}.
\end{equation*}
The ground state corresponds to $k=1$ and thus we get that the dynamically generated mass is related to the coupling constant as required by the Miransky scaling  
\begin{equation} \label{E25}
    \frac{m}{\Lambda} = \exp\left[-\frac{\pi}{\sqrt{\frac{\alpha}{\alpha_c}-1}}-2\ln 2\right].
\end{equation}

From all the previous discussion, it is well established that in QED, the mass function displays an infinite-order (conformal) phase transition and that the dynamically generated mass  follows the general law~\cite{Fomin:1976af,Miransky:1985gbh},
\bea
\label{gM}
    \frac{m}{\Lambda} = \exp\left[-\frac{A}{\sqrt{\frac{\alpha}{\alpha_c}-1}}-B\right],
\eea
where the constants $A$, $B$, and $C$ are determined based on the chosen truncation~\cite{Kizilersu:2014ela}. In the following, we present an analysis of the linearized gap equation in the form of a Schrödinger-like equation with an effective potential barrier (well) and address the problem of bound states.

\section{Gap equation as a Schrödinger-like equation}
\label{S-EQ}
\begin{figure}
    \centering
    \includegraphics[scale=0.45]{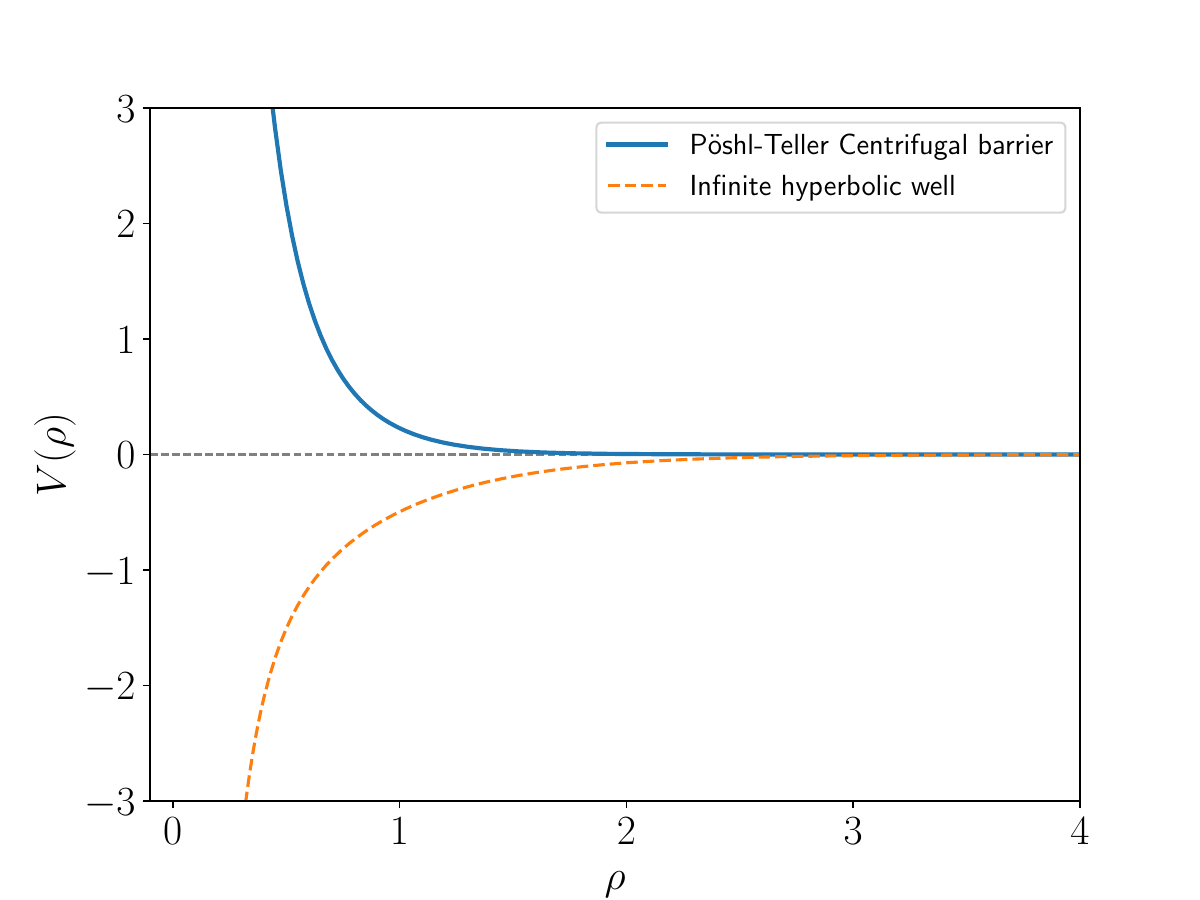}
    \caption{Plot of the Pöschl-Teller centrifugal barrier in equation \eqref{E5} and the infinite hyperbolic well in equation \eqref{Sch1} with $\hat{\alpha} = 1$. We can observe that in principle, these potentials do not support bound states.}
    \label{fig:PT}
\end{figure}

Upon taking the change of variable $x=\sinh^{2}\rho$, and taking $\Sigma(\rho)=\sinh^{-\frac{3}{2}}\rho\cosh^{\frac{1}{2}}\rho\Phi(\rho)$, the differential equation \eqref{hE2} can be written as
\begin{equation}\label{Sch1}
    -\frac{d^2\Phi}{d\rho^2}+\left[1+\frac{3}{\sinh^22\rho}-\hat{\alpha}\coth^2\rho\right]\Phi = 0.
\end{equation}
The potential term is an infinite hyperbolic well potential that, in principle, does not support bound states; see Fig.~\ref{fig:PT}. The general solution for the preliminary differential equation is
\begin{equation}
    \begin{aligned}
         \Phi(\rho) =& \sinh^{3/2}{\rho}\cosh^{-1/2}{\rho}\\
         &\times\left(C_+\sinh^{-1+i\tau}{\rho}+C_-\sinh^{-1-i\tau}{\rho}\right),\label{phihard}
    \end{aligned}
\end{equation}
which can be straightforward obtained from the solution in~\eqref{hE5}. Note that this solution satisfies the Schrödinger-like equation \eqref{Sch1} independently of the value of $\alpha$. Furthermore, the solution $\Phi(\rho)$ fulfills BC obtained from the IR and UV conditions in \eqref{hE3}, by applying the transformation and change of variable used to go from equation \eqref{hE2} to equation \eqref{Sch1}. Its asymptotic behavior when $\sinh{\rho}\rightarrow \kappa$ is given by  
\begin{equation}
    \lim_{\sinh{\rho}\rightarrow\kappa}\Phi(\rho) = \kappa^{3/2}(1+\kappa^2)^{-1/4}[C_+\kappa^{-1+i\tau}+C_-\kappa^{-1-i\tau}].
\end{equation}
From the IR condition, the factor in square brackets is a constant; thus, if $\kappa$ is  small enough, {\it i.e.,} $\kappa\rightarrow 0$, then $\Phi(\rho)\rightarrow 0$ when $\rho\rightarrow\sinh\rho\rightarrow 0$. Similarly, the behavior of $\Phi(\rho)$ when $\sinh{\rho}\rightarrow\Lambda$ is
\begin{equation}
    \begin{aligned}
        \lim_{\sinh\rho\rightarrow\Lambda}\Phi(\rho) =& \Lambda^{-1/2}(1+\Lambda^2)^{-1/4}\\
        &\times\Lambda^{2}[C_+\Lambda^{-1+i\tau}+C_-\Lambda^{-1-i\tau}].
    \end{aligned}
\end{equation}
From the UV condition, the factor $\Lambda^{2}[C_+\Lambda^{-1+i\tau}+C_-\Lambda^{-1-i\tau}]$ is constant; if $\Lambda$ is large  enough, that is, $\Lambda\rightarrow\infty$, then $\Phi(\rho)\rightarrow 0$ when $\rho\rightarrow\sinh\rho\rightarrow\infty$. Both assumptions are satisfied by the solutions that are ruled by the Miransky scaling law emerging from equation~\eqref{hardlin} 
and hence the function $\Phi(\rho)$ in \eqref{phihard} turns out to be the eigenfunction associated to a zero mode of the Schrödinger-like equation~\eqref{Sch1}.

Now, with the change of variable $z=-\sinh^{2}\rho$, and taking $\Sigma(\rho)=\sinh^{-\frac{3}{2}}\rho\cosh^{\frac{1}{2}}\rho\Phi(\rho)$, the differential gap equation~\eqref{E4} can be transformed to the Schrödinger-like form
\begin{equation} \label{E5}
    -\frac{d^2\Phi}{d\rho^2} + \frac{3}{\sinh^2 2\rho}\Phi = - 4\sigma^2\Phi.
\end{equation}
We must mention, the potential term in this differential equation corresponds to a centrifugal-like term that usually appears in the Pöschl-Teller potential~\cite{Poschl:1933zz}. In Fig.~\ref{fig:PT} we plot such potential, which is a infinite potential barrier. Notice that it is a monotonically decaying function of the coordinate $\rho$ and hence, at first glance, it would seem not to support bound states.

We determine the two solutions for the Schrödinger-like equation~\eqref{E5}, which are given by
\begin{align}
    \ \nonumber \Phi_{3/2}(\rho)&=\left(\sinh^{2}\rho\cosh^{2}\rho\right)^{3/4}\\ & \times F \left(\frac{3}{2}+\sigma,\frac{3}{2}-\sigma;2;-\sinh^{2}\rho\right),\label{E18}\\
    \ \nonumber \Phi_{1/2}(\rho) &= \left(\sinh^{2}\rho\right)^{3/4}\left(\cosh^{2}\rho\right)^{-1/4}\\
    &\times F\left(\frac{1}{2}+\sigma,\frac{1}{2}-\sigma;2;-\sinh^{2}\rho\right). \label{E19}
\end{align}
Furthermore, recalling  that the mass function is $\Sigma(\rho)=\sinh^{-3/2}\rho\cosh^{1/2}\rho\Phi(\rho)$, the solutions of the differential gap equation~\eqref{E4} in the original variables become
\begin{align}
    &\Sigma_{3/2}(z) = (1-z)F\left(\frac{3}{2}+\sigma,\frac{3}{2}-\sigma;2;z\right),\label{E20}\\
    &\Sigma_{1/2}(z) = F\left(\frac{1}{2}+\sigma,\frac{1}{2}-\sigma;2;z\right).\label{E21}
\end{align}
On the other hand, the solution $\Sigma_{3/2}(z)$ can be transformed to
\begin{equation}
    \Sigma_{3/2}(z)=F\left(\frac{1}{2}-\sigma,\frac{1}{2}+\sigma;2;z\right), \label{E22}
\end{equation}
where we used the connection formula 
\begin{equation}
F\left(a,b;c;z\right)=(1-z)^{c-a-b}F\left(c-a,c-b;c;z\right).
\end{equation}
Thus, the solutions to the Schrödinger-like equation \eqref{E5} are linked to the mass function of the differential gap equation \eqref{E4}.

\begin{figure}
    \centering
    \includegraphics[width=0.9\columnwidth]{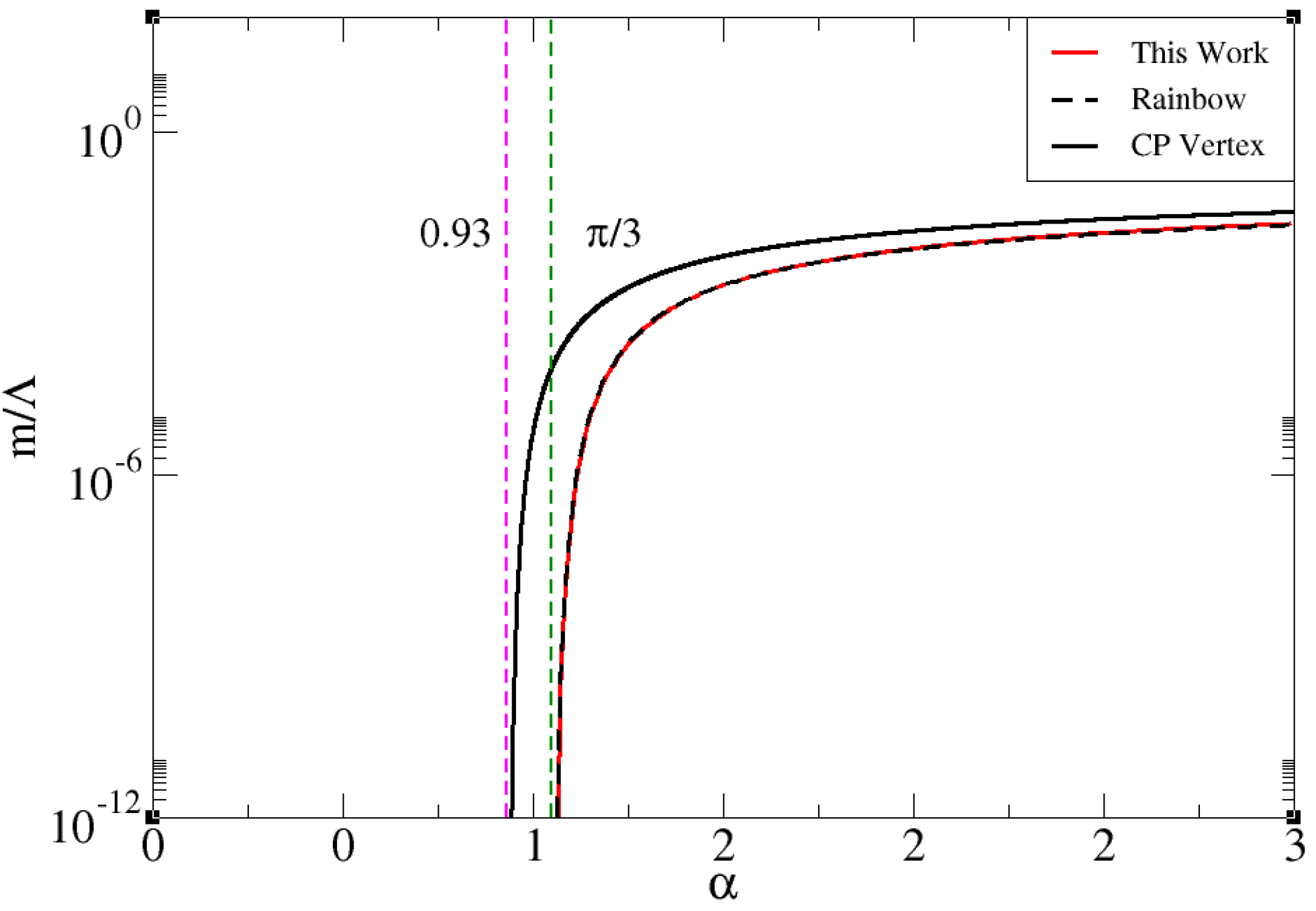}
    \caption{The dynamically generated mass in the rainbow approximation and with the CP-vertex as a function of the coupling. The critical coupling for the rainbow approximation is $\alpha_c=\pi/3$, while for the CP-vertex it is $\alpha_c=0.93$. As shown in the graph, the dynamical mass is generated from this critical value of  $\alpha_c$. 
    All curves were generated in the Landau gauge.}
    \label{fig:scalinglaws}
\end{figure}

\begin{table}[ht]
\caption{\label{scale-table}
Our result for equation~\eqref{gM} compared with other results. It is noteworthy that the critical coupling using the Curtis-Pennington (CP) vertex~\cite{Bloch} is $\alpha_c = 0.93$, which has a percentage difference of approximately 11\% from the value of $\pi/3$.}. 
 \renewcommand{\arraystretch}{1.6} %
\begin{tabular}{@{\extracolsep{0.5 cm}}||ccc||}
\hline \hline
 & A & B\\
 This Work & $\pi$ &  $2\ln 2 \sim 1.39$\\
 Rainbow   & 0.98$\pi$ & 1.07 $\ln 4 \sim 1.48$ \\
 CP Vertex & 0.93$\pi$ & 1.26 \\
 \hline \hline
 \end{tabular}
 \end{table}
As shown in Table \ref{scale-table} and Fig. \ref{fig:scalinglaws}, our results are consistent with other approximations, such as the Curtis-Pennington vertex~\cite{Curtis:1993py}. This explicit numerical and theoretical study backs up the conclusions made through a bifurcation analysis~\cite{Bashir:1994az} which has also been used to explore a critical number of flavors in the theory~\cite{Bashir:2011ij}.

\begin{figure}
    \centering
    \includegraphics[width=0.9\columnwidth]{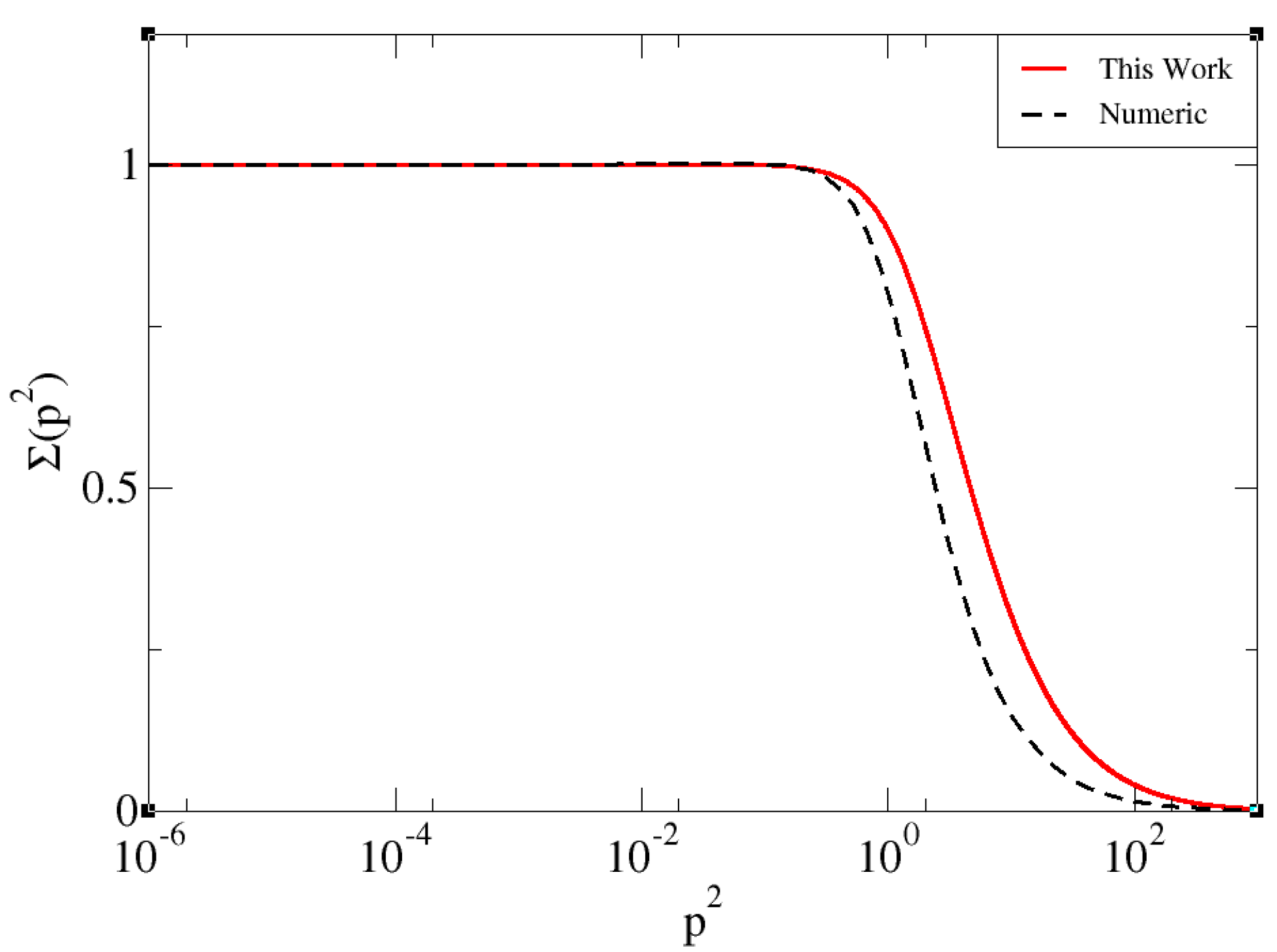}
    \caption{$\Sigma(p^2)$ mass function in equation (\ref{E22}), compared to the numerical solution of equation (\ref{E1}). As can be seen, the graphs are almost identical in the infrared. }
    \label{Sigma}
\end{figure}

In Fig. \ref{Sigma}, we present the results of equation~\eqref{E22}, compared with the numerical solution of the gap equation~\eqref{E1}.
Below we discuss and interpret de above result as a zero mode of the Schrödinger-like equation~\eqref{E5}. 

\begin{figure}
    \centering
    \includegraphics[scale=0.45]{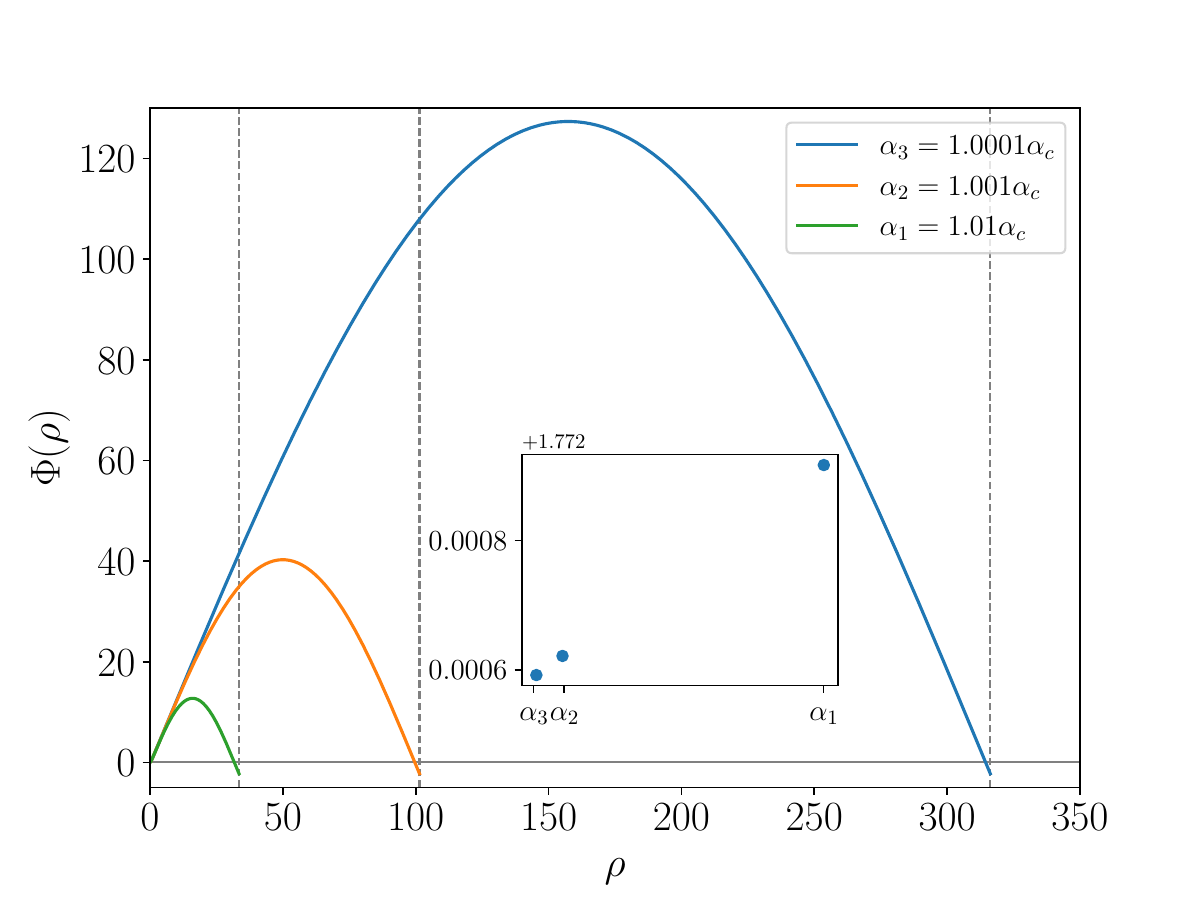}
    \caption{The solutions $\Phi_{3/2}(\rho)$ and $\Phi_{1/2}(\rho)$ in Equations \eqref{E18} and \eqref{E19}, respectively. Since both functions generated the same plot for each coupling parameter value, we only show one of them. The dashed gray vertical lines are the corresponding values of the UV cut-off for each coupling parameter value. And the small inside chart shows the difference between the value, where the eigenfunctions vanish, and the UV cut-off.}
    \label{fig2}
\end{figure}

As stated earlier, the solutions $\Phi_{3/2}(\rho)$ and $\Phi_{1/2}(\rho)$ in equations~\eqref{E18} and~\eqref{E19}, respectively, are associated to a eigenvalue $-4\sigma^{2}$. Furthermore, they satisfy asymptotic boundary conditions. Indeed, analyzing the asymptotic behavior of $\Phi_{1/2}(\rho)$ when $\rho\rightarrow 0$, we have that
\begin{align}
    \lim_{\rho\rightarrow 0} \Phi_{1/2}(\rho) &= \left(\sinh^{2}0\right)^{3/4}\left(\cosh^{2}0\right)^{-1/4}\\ 
    & \times F\left(\frac{1}{2}+\sigma,\frac{1}{2}-\sigma;2;-\sinh^{2}0\right) = 0.
\end{align}
Note this condition can be translated into the IR condition for the generated mass solution, using the transformation and change of variable used to go from equation \eqref{E4} to equation \eqref{E5}. Moreover, looking at the asymptotic behavior when $\rho\rightarrow\infty$, we arrive at
\begin{align}
    \lim_{\rho\rightarrow\infty} \Phi_{1/2}(\rho) &= \Gamma(2)\sqrt{\frac{\sinh\rho}{\cosh\rho}}\\
    &\times \Bigg[\frac{\Gamma(-2\sigma)}{(\frac{1}{2}-\sigma)\Gamma^{2}(\frac{1}{2}-\sigma)}\left(\sinh^{2}\rho\right)^{-\sigma}\\
    &+\frac{\Gamma(2\sigma)}{(\frac{1}{2}+\sigma)\Gamma^{2}(\frac{1}{2}+\sigma)}\left(\sinh^{2}\rho\right)^{\sigma}\Bigg].
\end{align}
We can observe that $\Phi_{1/2}(\rho) \rightarrow 0$ if the factor in square brackets vanishes. For the solutions following the Miransky scaling law, the coupling parameter $\alpha\sim\alpha_{c}$, and we can write $\sigma=i\tau/2$, with $\tau=\sqrt{\alpha/\alpha_c-1}$ a small real parameter. {\it i.e.}, $\tau\ll 1$. Therefore, the factor in square brackets vanishes in the limit when $\rho\rightarrow\infty$ as long as
\begin{equation}
    (\sinh^{2}\rho)^{i\tau} = -e^{i2\tau}\frac{\Gamma(-i\tau)\Gamma^{2}(\frac{1+i\tau}{2})}{\Gamma(i\tau)\Gamma^{2}(\frac{1-i\tau}{2})}.
\end{equation}
Recalling that $x/m^{2}=\sinh^{2}\rho$, the UV condition leads us to the Miransky scaling law for the dynamical mass with the coupling constant for small values of $\tau$. As a direct consequence of imposing to the generated mass solution $\Sigma_{1/2}(\rho)$ the UV condition, we have that $\Phi_{1/2}(\rho)\rightarrow 0$ when $\rho\rightarrow\infty$. The solution $\Phi_{3/2}(\rho)$ has the same asymptotic behaviors in the both limits (IR and UV). Hence, we can translate the boundary conditions in equation ~\eqref{E3} into asymptotic boundary conditions for the Schrödinger-like eqs.~\eqref{E5}, which are fulfilled by the solutions in equations~\eqref{E18} 
and~\eqref{E19}. 
Fig.~\ref{fig2} displays such functions for various values of the coupling constant. Both functions generate the same curve, thus, we only display one for each case. The dashed gray vertical lines are the corresponding values of the UV cut-off for each value of $\alpha$. In the inset of Fig.~\ref{fig2}, we show the parameter $\rho_0$ at which $\Phi_{1/2}(\rho_0)=0$ and the UV cut-off is displayed; we can observe that as the coupling constant approaches the critical value, the difference between the two slowly decreases and asymptotically vanishes. 
Thus, the Miransky scaling law leads us to $m/\Lambda\rightarrow 0$, {\it i.e.}, the UV cut-off $\Lambda$ is large enough to consider that the function $\Phi_{1/2}(\rho)$ is a nodeless eigenfunction associated to a bound state with 0 eigenvalue, a zero mode. 

It is worth mentioning that the differential gap equation for QED$_3$ \eqref{gap3} and its corresponding BC \eqref{BC3} have a form similar to the respective counterparts in QED$_4$, see equations \eqref{hE2} and \eqref{hE3}. Thus, through an analogous reasoning,  it is straightforward to show that the power law solution~\eqref{sol3} transforms to an eigenfunction with 0 eigenvalue of the corresponding Schrödinger-like equation, when the parameter $N_f\rightarrow N_f^c$. 

%

\section{Summary and outlook}
\label{Summary}

In this article, we revisited the SDE for the electron propagator in QED in three- and four-space--time dimensions. Under minimalistic assumptions on the interaction kernel, our analysis  allowed to explore how the non-perturbative dynamics associated with chiral symmetry breaking manifest across different dimensional settings. This framework provides valuable insight into the mechanisms of mass generation and the role of gauge interactions beyond perturbation theory.

 Consistent with bifurcation analysis, the linearized integral equation is transformed into a differential equation for the mass function, which must satisfy boundary conditions at both high and low momenta. Chiral symmetry and the dynamical generation of fermion mass are observed when the coupling constant $\alpha$ exceeds a critical value $\alpha_c$. Near this critical point, a conformal phase transition associated to the break down of chiral symmetry such that the dynamically generated mass exhibits Miransky scaling as the coupling increases. These findings are consistent with established results in QED. The distinctive aspect of our approach lies in the transformation of the hypergeometric equation into a Schrödinger-like equation featuring a Pöschl-Teller centrifugal barrier, which does not admit bound states. However, this equation can be solved using localized, square-integrable states of the mass function, with vanishing eigenvalues of the auxiliary Hamiltonian operator. These zero modes  provide a natural explanation for the quantization rule associated with the Miransky scaling law (MSL).Excited modes are connected with the Sturm-Liouville property of the gap equation, as discussed in~\cite{chang,multiple}. The criticality of the coupling is also linked to a critical condition for the number of active fermion families $N_f$ in QED, as discussed in \cite{Bashir:2011ij}. Notice, however, that the scaling of the dynamical mass differs from the Miransky scaling discussed in this article.

In the low-dimensional version of QED, within the $1/N_f$ approximation—relevant for condensed matter systems such as graphene—the gap equation admits solutions up to a critical value of $N_f$, beyond which chiral symmetry is restored. These solutions also exhibit Miransky scaling, and our formalism describes them as zero modes.

In QCD, with a specific choice of the quark-gluon vertex and a model for the running coupling, the gap equation becomes a numerical problem that generally lacks analytical solutions. Although a critical value of the coupling may exist for the dynamical generation of quark masses, the strong coupling regime at low energies places this critical scenario well beyond the scale of hadron physics. Nonetheless, it is plausible that an appropriate linearization of the gap equation near a theoretical critical point could lead to a more complex form of the auxiliary Schrödinger-like equation. Whether such a linearized equation supports zero modes is currently under investigation. Properties of bound states within a Lipmann-Schwinger framework are also considered. Results will be presented in future work.

We dedicate this work to the loving memory of Prof. Vladimir Miransky, who inspired us to explore non-perturbative phenomena in quantum field theory.

\section*{Acknowledgements}
The authors gratefully acknowledge the invaluable comments and suggestions provided by the referee.
LXGG, AA and JDGM acknowledge the {\em Secretaría de Ciencia, Humanidades, Tecnología e Inovación} (SECIHTI) for support through Project CBF2023-2024-268, Hadronic Physics at JLab: Deciphering the Internal Structure of Mesons and Baryons, from the 2023-2024 frontier science call. LXGG acknowledges SECIHTI for the support provided to her through the {\em Investigadores e Investigadoras por México} SECIHTI program. AR acknowledges support from FORDECYT-PRONACES/61533/2020 and CIC-UMSNH  under grant 18371. JDGM acknowledges support from SECIHTI for the postdoctoral fellowship with CVU number 712251. 

\bibliography{References}
\bibliographystyle{ieeetr}
\end{document}